\documentclass[12pt]{article}
\usepackage{amsmath}
\usepackage{amsthm}
\usepackage{amssymb}
\usepackage{latexsym}
\usepackage{subeqnarray}
\usepackage{graphicx}
\usepackage[cp1251]{inputenc}
\inputencoding{cp1251}
\usepackage[T2B]{fontenc}
\usepackage[russian,english]{babel}
\captionsrussian

\newcommand{\bit}{\begin{itemize}}
\newcommand{\eit}{\end{itemize}}
\newcommand{\nc}{\newcommand*}

\nc{\nn}{\nonumber}
\def\ri{{\rm i}}

\begin{document}

\begin{center}
{\large\bf Damaskinsky E.V. \footnote{Math. Dept. Military Engineering Institute. VI(IT).
Saint Petersburg, evd@pdmi.ras.ru},\, Sokolov M.A \footnote{St. Petersburg state Peter The Great
Polytechnical University; masokolov@gmail.com}}
\bigskip

{\Large\bf The generating function of bivariate Chebyshev polynomials
associated with the Lie algebra $G_2$ \footnote{The work is supported by RFBR under the grant 15-01-03148}}
\end{center}
\medskip

\begin{flushright}
{\it Dedicated to the memory of our dear friend Peter Kulish} \\
\end{flushright}
\medskip

\begin{quote}
The generating function of the second kind bivariate Chebyshev polynomials associated with the simple Lie algebra $G_2$ is constructed by the method proposed in \cite{DKS} and \cite{DKS1}.
\end{quote}
\bigskip

\section{Introduction}
The present work finishes the series of the articles (\cite{DKS}, \cite{DKS1}, \cite{MAS}) which were initiated by P. Kulish.
In these works we propose the method of constructing of generating functions for Chebyshev polynomials
in several variables (of the first and second kinds), associated with simple Lie algebras.
The method was tested on examples of polynomials related to Lie algebras $A_2$, $C_2$, $A_3$ and
$G_2$ (in the last case only for the polynomials of the first kind).
In this work we consider the case of bivariate Chebyshev polynomials of the second kind associated
with the algebra $G_2$, thus complete the consideration of polynomials related with
simple Lie algebras of rank 2.

Attention of P.Kulish in this subject arose during the preparation of his report at the conference
in Nankai \cite{Nankai}. During the discussion of the report with one of the authors (EVD),
it was found that the Chebyshev polynomials of the 2-nd kind of one or two variables naturally  arise
when considering the integrable spin chains by the quantum inverse scattering method.

It was assumed that this topic will be discussed in detail in further studies, however, the illness and death of
P. Kulish prevented the implementation of this plan.

In this paper we construct generating function of bivariate Chebyshev polynomials of the second kind
associated with the simple Lie algebra $G_2$ and make a few comments regarding this construction.
We will consider the most natural variant of generalization of Chebyshev polynomials to the case of
several variables suggested in the work of Koornwinder \cite{K1} (see also \cite{Suet2}, \cite{K2}).

\section{Description of the method}

Chebyshev polynomials in several variables are a natural generalization of the classical
Chebyshev polynomials of one variable (see for example \cite{Suet1}, \cite{Ri}). They have
application in various areas of mathematics (in the theory of approximations \cite{RM-K},
in the linear algebra \cite{SS}, \cite{AS}, in the theory of representations \cite{KLP}-\cite{L})
and in physics (\cite{GR}-\cite{BD1}).

The classical Chebyshev polynomials of the first kind $T_n(x)$ are defined by the formula
\begin{equation}
\label{1-1}
T_n(x)=T_n(\cos{\phi})=\cos{n\phi},\quad n\ge 0,
\end{equation}
where $\phi=\arccos\,x$.
Polynomials of the second kind $U_n(x)$ is defined as
\begin{equation}
\label{1-2}
U_n(x) = \frac{\sin{(n+1)\phi}}{\sin{\phi}},\quad n\ge 0.
\end{equation}
Both polynomials satisfy the well known three-term recurrence relation
\begin{equation}
\label{1-3}
T_{n+1}(x)=2x\,T_n(x)-T_{n-1}(x),\quad n\ge 1,
\end{equation}
but with different initial conditions
$$
T_0(x)=1,\quad T_1(x)=x,\quad U_0(x)=1, \quad U_1(x)= 2x.
$$

Chebyshev polynomials of several variables can be determined
for any simple Lie algebra in the following way. Let $L$ be the simple Lie algebra with
reduced root system  $R$ consisting of vectors in $d$-dimensional Euclidean space $E^d$ with
the scalar product $ (.,.) $.

The root system $R$ is completely determined by the basis of simple roots $\alpha_i,\, i=1,\ldots,d$ and
finite reflection group $W(R)$ --- the Weyl group. Generating elements of the Weyl group
$w_i,\, i=1,\ldots,d$ acts on the simple roots by the rule $w_i\,\alpha_i=-\alpha_i$.
The root system $R$ is closed under the action of the Weyl group. For any vector $x\in E^d$ elements
$w$ of the group $W(R)$ acting according to the formula
\begin{equation}\label{1-4}
w_i\,x=x-\frac{2(x,\alpha_i)}{(\alpha_i,\alpha_i)}\alpha_i.
\end{equation}
To each root $\alpha\in R$ correspond associated co-root
$$
\alpha^{\vee}=\frac{2\alpha}{(\alpha,\alpha)}.
$$
For the basis of simple co-roots $\alpha^{\vee}_i, \, i=1,\ldots,d$ one can define the dual basis
of the fundamental weights $\lambda_i, \, i=1,\ldots,d$
$$
(\lambda_i,\alpha^{\vee}_j)=\delta_{ij}
$$
(the dual space ${E^d}^*$ is identified with $E^d$). The bases of the roots and weights are related
by the linear transformation
\begin{equation}\label{1-5}
\alpha_i=C_{ij}\lambda_j, \quad C_{ij}=\frac{2(\alpha_i,\alpha_j)}{(\alpha_j,\alpha_j)},
\end{equation}
where $C$ is the Cartan matrix of the algebra $L$. Recall that the bases of roots and weights are not
orthonormal.

The Chebyshev polynomials of the first kind associated with the Lie algebra $L$ is defined via
$W$-invariant functions
\begin{equation}\label{1-6}
\Phi_{\bf n}({\boldsymbol{\phi}}) =
\sum\limits_{w\in {\mbox{\footnotesize W}}}e^{2\pi\ri (\emph{w}\,{\bf n},{\boldsymbol{\phi}})},
\end{equation}
where vector $\bf n$ with nonnegative integer components expressed in the basis of
the fundamental weights $\{\lambda_i\}$ and ${\boldsymbol{\phi}}$ -- in the dual basis of
co-roots $\{\alpha^\vee_i\}$
$$
{\bf n} \equiv (n_1,\,...,\,n_d) =\sum_{i=1}^d\,n_i\lambda_i \quad n_i\in N, \quad  {\boldsymbol{\phi}}=
\sum_{i=1}^d\,\phi_i\alpha_i^\vee \quad \phi_i\in [0,1).
$$
Defining new variables (generalized cosines) $x_i,\,i=1,..,d$ by relations
\begin{equation}\label{1-7}
x_i\,=\,\Phi_{{\bf e}_i}({\boldsymbol{\phi}}),\quad {\bf e}_i
= (\overbrace{0,..,0}^{i-1},1,\overbrace{0,..,0}^{d-i}).
\end{equation}
and recording function (\ref{1-6}) in terms of $x_i$ we come to the (non-normalized)
Chebyshev polynomials of the first kind in $d$ variables. The fact that this procedure
is possible for any simple Lie algebra, shown in the works \cite{K1}, \cite{H} - \cite{LU}.

The function $\Phi_{\bf n}({\boldsymbol{\phi}})$ has many useful properties, including the following
"rule of multiplication"
\begin{equation}
\label{1-8}
\Phi_{\bf k}\Phi_{\bf s}=\sum\limits_{w\in W}\Phi_{w\,{\bf k}+{\bf s}},\quad {\bf k}=(k_1,\,...,\,k_d),\quad
{\bf s}=(s_1,\,...,\,s_d),
\end{equation}
which allows to find the recurrence relation for polynomials in several variables.

The Chebyshev polynomials of the second kind associated with the Lie algebra $L$, are
determined by the relations
\begin{equation}\label{1-9}
U_{\bf n}({\boldsymbol{\phi}})=\frac{\sum\limits_{w\in {\mbox{\footnotesize W}}}\det{w}\,\,
e^{2\pi\ri (\emph{w}\,({\bf n}+\boldsymbol{\rho}),{\boldsymbol{\phi}})}}
{\sum\limits_{w\in {\mbox{\footnotesize W}}}\det{w}\,\,
e^{2\pi\ri (\emph{w}\,{\boldsymbol{\rho}},{\boldsymbol{\phi}})}},
\end{equation}
where $\det{w}=(-1)^{\ell\,(\textit{w})}$ and ${\ell}(\textit{w})$ is the minimum number of
generators $w_i$ of the Weyl group $w_i$ needed to represent the element of $w$
by product of generators from $w_i$.
In (\ref{1-9}) $\boldsymbol{\rho}$ is the Weyl vector equal to
the sum of positive roots. With the help of relations (\ref{1-5}) this vector can be
described in terms of the fundamental weights. As in the case of polynomials of the first kind,
one can define new variables $x_i = U_{{\bf e}_i}({\boldsymbol{\phi}})$,
in terms of which can be described function (\ref{1-9}) for any non-negative $n_i$
from multi-index ${\bf n}=(n_1,\,...,\,n_d)$. The function $U_{\bf n}({\boldsymbol{\phi}})$,
presented in terms of $x_i$, gives the second kind Chebyshev polynomials in several variables.

In practice describing of the functions $\Phi_{\bf n}({\boldsymbol{\phi}})$ (\ref{1-6}) and
$U_{\bf n}({\boldsymbol{\phi}})$ (\ref{1-9}) (with given multi-index
${\bf n}=(n_1,\,...,\,n_d)$) in terms of the variables $x_i$ is quite cumbersome task.
So obviously, one need an effective method of direct computation of Chebyshev polynomials in
several variables. In the works \cite{DKS}, \cite{DKS1} we propose a new method for calculation
of generating functions for Chebyshev polynomials of any kind associated with any simple Lie algebra.
Let us describe it briefly.

Let's start with the function (\ref{1-6}) that defines the Chebyshev polynomials of the first kind.
Because components of the vector ${\bf n}$ are all non-negative numbers, the scalar product in
the functions $\Phi_{\bf n}({\boldsymbol{\phi}})$ can be represented in the form
$(\textit{w}\,{\bf n},{\boldsymbol{\phi}}) = \sum_k w (\lambda_k,{\boldsymbol{\phi}})n_k$,
and the function $\Phi_{\bf n}({\boldsymbol{\phi}})$ can be expressed as the following products
\begin{equation}\label{1-10}
\Phi_{\bf n} = \sum\limits_{w\in {\mbox{\footnotesize W}}}\prod_k\left(e^{2\pi\ri
(w\lambda_k,{\boldsymbol{\phi}})}\right)^{n_k} = {\rm tr}\left(\prod_kM_k^{n_k}\right).
\end{equation}
Here $M_k$ --- diagonal matrices, nonzero elements of which are the components of the sum (\ref{1-6})
\begin{equation}\label{1-11}
M_k={\rm diag}
(e^{2\pi\ri (w_1\lambda_k,{\boldsymbol{\phi}})},
e^{2\pi\ri (w_2\lambda_k,{\boldsymbol{\phi}})},....,
e^{2\pi\ri (w_{|W|}\lambda_k,{\boldsymbol{\phi}})}),
\end{equation}
where $w_i$ --- elements of the Weyl group $W$, and $|W|$ is the number of elements of this group.

We introduce a new matrix $R_k=(I_{|W|}-p_kM_k)^{-1}$, where $I_{|W|}$ is $|W|$-dimensional
identity matrix and $p_k$ are arbitrary real parameters. In this notation
the function $\Phi_{\bf n}$ has the form
\begin{equation}\label{1-12}
\Phi_{{\bf n}}({\boldsymbol{\phi}})=\Phi_{{n_1,..,n_d}}({\boldsymbol{\phi}}) =
\frac1{n_1!..n_d!}\left.\frac{\partial ^{n_1+..+n_d}}{\partial ^{n_1}p_1..\partial ^{n_d}p_n}
\left({\rm tr}(R_{p_1}..R_{p_d})
\right)\right|_{p_1= .. =p_d = 0}.
\end{equation}
The simple structure of matrices $R_{p_k}$ allows to represent the function ${\rm tr}(R_{p_1}..R_{p_d})$
in terms of the variables defined  above (\ref{1-7})
$$
x_i\,=\,\Phi_{{\bf e}_i}({\boldsymbol{\phi}}),\quad {\bf e}_i =
(\overbrace{0,..,0}^{i-1},1,\overbrace{0,..,0}^{d-i}).
$$
The coefficients $\Phi_{{n_1,..,n_d}}({\boldsymbol{\phi}})$  of the function
\begin{equation}\label{1-13}
F_{p_1,..,p_d}^I ={\rm tr}(R_{p_1}..R_{p_d}) =
\sum\limits_{n_1..n_d\ge 0}\Phi_{{n_1,..,n_d}}(x_i)p_1^{n_1}..p_d^{n_d}
\end{equation}
also can be expressed in terms of $x_i$ and are therefore the (non-normalized) Chebyshev polynomials
of the first kind in $d$ variables. The function $F_{p_1,..,p_d}^I$ is the generating
function of these polynomials.

The computation of the generating functions for the second kind Chebyshev polynomials in
$d$ variables required only a small modification of the method described above. Consider a function
$U_{\bf n}({\boldsymbol{\phi}})$ that determines these polynomials in the form
\begin{equation}\label{1-14}
U_{\bf n}({\boldsymbol{\phi}})=
\frac{\Phi_{\bf n+\boldsymbol{\rho}}^{as}}{\Phi_{\boldsymbol{\rho}}^{as}}=
\frac{\Phi_{\bf n+\boldsymbol{\rho}}^{as+}-
\Phi_{\bf n+\boldsymbol{\rho}}^{as-}}{\Phi_{\boldsymbol{\rho}}^{as+}-
\Phi_{\boldsymbol{\rho}}^{as-}},
\end{equation}
where
$$
\Phi_{\bf k}^{as\pm} = \sum\limits_{w\in {\mbox{\footnotesize W}},\,
\det{w}=\pm 1}\,e^{2\pi\ri (\emph{w}\,{\bf k},{\boldsymbol{\phi}})}.
$$
Further calculations, repeat the considerations given above (\ref{1-10}) - (\ref{1-13}) for each function
$\Phi_{\bf k}^{as\pm}$ separately. In the result  the generating function for the polynomials of the
second kind takes the form
\begin{equation}\label{1-16}
F_{p_1,..,p_d}^{II} = \frac{{\rm tr}(R^+_{p_1}..R^+_{p_d}-R^-_{p_1}..R^-_{p_d})}{\Phi_{\boldsymbol{\rho}}^{as}},
\end{equation}
where the matrix $R^{\pm}$ are expressed in terms of the functions $\Phi_{\bf k}^{as\pm}$.
Since the Weyl vector in the space of the fundamental weights is given by the formula
${\boldsymbol{\rho}} = \sum_{i=1}^d{\lambda_i}$,
the multiplier $\Phi_{\boldsymbol{\rho}}^{as}$ at the denominator of (\ref{1-14}), is calculated as
\begin{equation}\label{1-17}
\Phi_{\boldsymbol{\rho}}^{as} =
\left.\frac{\partial ^{d}}
{\partial p_1..\partial {p_d}}
\left({\rm tr}(R^+_{p_1}..R^+_{p_d}-R^-_{p_1}..R^-_{p_d})
\right)\right|_{p_1= .. =p_d = 0}.
\end{equation}
Thus the Chebyshev polynomials of the second kind are given by the expressions
\begin{equation}\label{1-18}
U_{\bf n} =
\frac1{(n_1+1)!..(n_d+1)!}\left.\frac{\partial ^{n_1+..+n_d+d}}{\partial ^{n_1+1}p_1..\partial ^{n_d+1}p_d}
\left(
\frac{{\rm tr}(R^+_{p_1}..R^+_{p_d}-R^-_{p_1}..R^-_{p_d})}{\Phi_{\boldsymbol{\rho}}^{as}}
\right)\right|_{p_1=\cdots=p_d = 0}.
\end{equation}

\section{The Chebyshev polynomials of the second kind associated with the Lie algebra $G_2$}

Using the above scheme, we calculate the generating function for bivariate Chebyshev polynomials
of the second kind, associated with the exceptional Lie algebra $G_2$. The root system of this algebra has
two fundamental roots $\alpha _1,\,\alpha_2$ and includes the following positive roots
$\alpha _1+\alpha_2,\,2\alpha _1+\alpha_2,\,3\alpha _1+\alpha_2,\,3\alpha _1+2\alpha_2$
along with their reflections.

Using the Cartan matrix
\begin{equation*}
C_{G_2}=\left(
\begin{array}{cc}
2 & -1 \\
-3 & 2 \\
\end{array}
\right)
\end{equation*}
and the formula (\ref{1-5}), we obtain the action of the generating elements
$w_1,w_2$ of the Weyl group $W(G_2)$ on the fundamental roots and fundamental weights
$$
w_1\alpha_1=-\alpha_1,\quad w_1\alpha_2=3\alpha_1+\alpha_2,\quad w_2\alpha_1=\alpha_1+\alpha_2,
\quad w_2\alpha_2=-\alpha_2.
$$
$$
w_1\lambda_1=\lambda_2-\lambda_1,\quad w_1\lambda_2=\lambda_2,\quad w_2\lambda_1=\lambda_1,
\quad w_2\lambda_2=2\lambda_1-\lambda_2.
$$
The expression of other elements of Weyl groups via generators looks as
$$w_3=w_1w_2,\, w_4=w_2w_1,\, w_5=w_2w_1w_2,\, w_6=w_1w_2w_1,\, w_7=(w_1w_2)^2,$$
$$\!w_8\!=\!(w_2w_1)^2,\, w_9\!=\!w_2(w_1w_2)^2,\, w_{10}\!=\!w_1(w_2w_1)^2,\,
w_{11}\!=\!(w_1w_2)^3,\, w_{0}\!=\!e,$$
and allows to extend the above relations to the whole group. Determinants of the
elements $w_1,w_2,w_5,w_6,w_9,w_{10}$ of the Weyl group is equal to $-1$,
the determinant of the remaining elements are equal to unit.

Using the above relations and the definition
$$
\Phi_{\bf k}^{as} = \sum\limits_{w\in {\mbox{\footnotesize W}}}\,\det{w}\,e^{2\pi\ri
(\emph{w}\,{\bf k},{\boldsymbol{\phi}})},
$$
we obtain
\begin{align*}
\Phi_{m,n}^{as} =\Phi_{m,n}^{as+}-\Phi_{m,n}^{as-}&\!=
          \!\left[e^{2\pi\ri (m\phi + n\psi)}\!+\!
            e^{2\pi\ri (m(-\phi +\psi) + n(-3\phi +2\psi))}\!+\!
            e^{2\pi\ri (m(2\phi -\psi) + n(3\phi -\psi))}\!+\right.\\
         &  \left.e^{-2\pi\ri (m\phi + n\psi)}\!+\!
            e^{-2\pi\ri (m(-\phi +\psi)\!+\!n(-3\phi +2\psi)0}\!+\!
            e^{-2\pi\ri (m(2\phi -\psi) + n(3\phi -\psi))}\right]\!-\\
         & \left[e^{2\pi\ri (m \phi +n(3\phi-\psi))}\!+\! e^{2\pi\ri (m(-\phi + \psi)+n\psi}\!+\!
            e^{2\pi\ri (m(2\phi-\psi) + n(3\phi -2\psi))}\!+\right.\\
         &  \left. e^{-2\pi\ri (m \phi +n(3\phi-\psi))}\!+\! e^{-2\pi\ri (m(-\phi + \psi)+n\psi)}\!+\!
            e^{-2\pi\ri (m(2\phi-\psi) + n(3\phi -2\psi))}\right],
\end{align*}
where the indices are marked by $m,n$, and angles by $\phi, \psi$.

We introduce the following four diagonal matrices
\begin{align*}
M_1^+ =& {\rm diag}(e^{2\pi\ri \phi},\, e^{2\pi\ri (-\phi +\psi)},\, e^{2\pi\ri (2\phi -\psi))},\,
         e^{-2\pi\ri (2\phi -\psi)},
        e^{-2\pi\ri (-\phi +\psi)},\,e^{-2\pi\ri \phi}),\\
M_2^+ =& {\rm diag}(e^{2\pi\ri \psi},\, e^{2\pi\ri (-3\phi +2\psi)},\, e^{2\pi\ri (3\phi -\psi)},\,
         e^{-2\pi\ri (3\phi -\psi)},
        e^{-2\pi\ri (-3\phi +2\psi)},\, e^{-2\pi\ri \psi}),\\
M_1^- =& {\rm diag}(e^{2\pi\ri (-\phi + \psi)},\, e^{2\pi\ri \phi},\, e^{2\pi\ri (2\phi -\psi)},\,
         e^{-2\pi\ri (2\phi -\psi)},
          e^{-2\pi\ri (-\phi + \psi)},\,e^{-2\pi\ri \phi}),\\
M_2^- =& {\rm diag}(e^{2\pi\ri \psi},\, e^{2\pi\ri (3\phi -\psi)},\, e^{2\pi\ri (3\phi -2\psi)},\,
         e^{-2\pi\ri (3\phi -2\psi)},
      e^{-2\pi\ri \psi},\,e^{-2\pi\ri (3\phi -\psi)}).
\nonumber
\end{align*}
Then for the function $\Phi_{m,n}^{as}$ we obtain the representation
\begin{equation*}
\Phi_{m,n}^{as} = {\rm tr}((M^+_1)^m(M^+_2)^n-(M^-_1)^m(M^-_2)^n).
\end{equation*}
The Weyl vector has the form
\begin{equation*}
\boldsymbol{\rho} = \frac{1}{2}(\alpha_1+\alpha_2) = \lambda_1+\lambda_2.
\end{equation*}
So, according to the definition given above, we introduce new variables
\begin{equation*}
x = U_{1,0}=\frac{\Phi_{2,1}^{as}}{\Phi_{1,1}^{as}},\quad y = U_{0,1} = \frac{\Phi_{1,2}^{as}}{\Phi_{1,1}^{as}}.
\end{equation*}
Taking into account the expression of singular element
\begin{align*}
{\Phi_{1,1}^{as}}&= e^{2\pi\ri (\phi +\psi)}-e^{2\pi\ri (-\phi +2\psi)}+e^{2\pi\ri (-4\phi +3\psi)}-
           e^{2\pi\ri (4\phi -\psi)}+e^{2\pi\ri (5\phi -2\psi)}-e^{2\pi\ri (5\phi -3\psi)}+\\
            &+ e^{2\pi\ri (-5\phi +2\psi)}-
           e^{2\pi\ri (-5\phi -3\psi)}
           +e^{2\pi\ri (4\phi -3\psi)}-e^{2\pi\ri (\phi -2\psi)}+e^{2\pi\ri (-\phi -\psi)}-
           e^{2\pi\ri (-4\phi +\psi)},
\end{align*}
after a simple calculation we obtain the following expressions for $x,y$
\begin{equation*}
x =  e^{-2\pi\ri\phi}+e^{2\pi\ri (\phi -\psi)}+e^{2\pi\ri (-2\phi +\psi)}
   +e^{2\pi\ri (2\phi -\psi)}+e^{2\pi\ri (-\phi +\psi)}+e^{2\pi\ri \phi}+1,
\end{equation*}
\begin{align*}
y &= e^{2\pi\ri (-3\phi +\psi)}+e^{2\pi\ri (-\psi)}+e^{2\pi\ri (3\phi -2\psi)}
    +e^{2\pi\ri (3\phi -\psi)}+e^{2\pi\ri (3\phi +2\psi)}+e^{2\pi\ri\psi}\\
  & +e^{2\pi\ri (-\phi)}+e^{2\pi\ri (\phi -\psi)}+e^{2\pi\ri (-2\phi +\psi)}
   +e^{2\pi\ri (2\phi -\psi)}+e^{2\pi\ri (-\phi +\psi)}+e^{2\pi\ri \phi}+2.
\end{align*}

Turning to the construction of a generating function, we introduce the matrix
\begin{equation*}
R_p^{\pm}=(I_6-pM_1^{\pm})^{-1},\quad R_q^{\pm}=(I_6-qM_2^{\pm})^{-1},
\end{equation*}
where $I_{6}$ is the identity matrix of order $6$, $p$ and $q$ real parameters, and the function
\begin{equation*}
{\Phi_{p,q}^{as}}(\phi,\psi) = {\rm tr}(R^+_{p}R^+_{q}-R^-_{p}R^-_{q}).
\end{equation*}
This function factorizes, and the multiplier is the singular element of ${\Phi_{1,1}^{as}}$.

Computing the ratio ${\Phi_{p,q}^{as}}/{\Phi_{1,1}^{as}}$ and representing it in terms of new
variables $x,y$, we come to the generating function for the Chebyshev polynomials of the second kind,
associated with the Lie algebra $G_2$
\begin{equation*}
F^{G_2}(p,q;x,y) = \frac{{\Phi_{p,q}^{as}}}{{\Phi_{1,1}^{as}}} =
(P_1P_2)^{-1}\left(\sum\limits_{i,j=0}^4K_{ij}p^iq^j\right).
\end{equation*}
In this expression, the polynomials $P_1,\,P_2$ are defined by the formulas
\begin{equation*}
P_1=1+(1-x)p+(y+1)p^2-(x^2-2y-1)p^3+(y+1)p^4+(1-x)p^5+p^6,
\end{equation*}
\begin{multline*}
P_2=1+(x-y+1)q+(x^3-3xy-2y-x+1)q^2-\\
        (y^2-2x^3+4xy+6y+x^2-2y+2x-1)q^3+\\
                  (x^3-3xy-2y-x+1)q^4+(x-y+1)q^5+q^6,
\end{multline*}
and non-zero coefficients $K_{ij}$ have the form
\begin{gather*}
K_{00}=1,\, K_{10}= 1,\,  K_{01}=x+1,\\
K_{02}=x+1,\, K_{11}=-x^2+x+y+2,\\
K_{03}=1,\, K_{12}=-x^2+x+1,\, K_{21}=y+1, \\
K_{31}=1-x,\, K_{13}=1-x,\, K_{22}=-x^2+xy+y+2x+1,\\
K_{41}=1,\, K_{23}=y+1,\, K_{32}= -x^2+x+1,\\
K_{33}=-x^2+x+y+2,\, K_{42}=x+1,\\
K_{34}=1,\, K_{43}=x+1,\, K_{44}=1.
\end{gather*}
Using the generating function
\begin{equation*}
U_{m,n}(x,y) = \frac1{m!n!}\left.\frac{\partial ^{m+n}}{\partial ^{m}\partial ^{n}}
\left(F^{G_2}(p,q;x,y)\right)\right|_{p=q=0},
\end{equation*}
we write out the first few Chebyshev polynomials of the second kind associated with the root
system of the Lie algebra $G_2$.

\begin{eqnarray*}
U_{0,0}&=& 1\\
U_{1,0}&=& x \\
U_{0,1}&=& y \\
U_{2,0}&=& x^2-x-y-1\\
U_{1,1}&=&-x^2+xy+y+1\\
U_{0,2}&=&-x^3+2xy+y^2+2x+y\\
U_{3,0}&=&-2xy-x-x^2-y+x^3\\
U_{2,1}&=&-y+x+x^2-y^2-x^3+x^2y\\
U_{1,2}&=&2x^2-x-1-x^4+x^2y+x^3+y^2+xy^2\\
U_{0,3}&=&-2x^3y+4xy^2+3y^2+4xy+2y-x^3-x^2+x^4-2x^2y+y^3\\
U_{4,0}&=& y^2+2y+x-3x^2y-x^2-x^3+x^4\\
U_{3,1}&=& 2x^3-2y^2-2x-2y+x^2y+x^3y-2xy^2-2xy-x^4+x^2\\
U_{2,2}&=&1+2x^3-2y^2-x-x^5-y^3-x^2y+2x^3y-2xy^2-4xy+x^2y^2+2x^4-4x^2\\
U_{1,3}&=& -4x^3+y^2+2x+2x^5+y^3+4x^2y-4x^3y+4xy^2+
        6xy+3x^2y^2-2x^4y+xy^3\\
        &&{}-2x^4+2x^2\\
U_{0,4}&=&-1+2x^3+4y^2-x-2y-x^5+5y^3+6x^2y-2x^3y+
        9xy^2+2xy-2x^4y-3x^3y^2\\
        &&{}+6xy^3-3x^4+3x^2+x^6+y^4.
\end{eqnarray*}
These polynomials coincide with the polynomials found from recurrence
relations in the work \cite{LU}

In the presented derivation of generating function is used only the definition of Chebyshev polynomials
and the Weyl character formula and absolutely not involved recurrence relations.
Since such relations provide several new features, we give in conclusion the following remarks.

The easiest way for obtaining the recurrence relations for Chebyshev polynomials of several variables
is to use (\ref{1-8})
\begin{equation*}
\Phi_{\bf k}\Phi_{\bf s} = \sum_{w\in W}\Phi_{w{\bf k}+{\bf s}},
\end{equation*}
where the summation in the right hand part applies to all elements of the Weyl group; ${\bf k}$ and ${\bf s}$
are vectors in the space of the fundamental weights with non-negative integer components
${\bf s} = (m,n)$. In the case of two variables, we assume that ${\bf k}$ in this space is equal to
${\bf k} = (1,0)$ firstly, and then to ${\bf k} = (0,1)$ and taking in to account the above formula
for action generators of the Weyl group on the fundamental weights, we obtain in a convenient normalization
\begin{align}
xU_{m,n} &= U_{m+1,n}+U_{m-1,n+1}+U_{m+2,n-1}+U_{m-2,n+1}+U_{m+1,n-1}+U_{m-1,n} \nn \\
yU_{m,n} &= U_{m,n+1}+U_{m+3,n-1}+U_{m-3,n+2}+U_{m+3,n-2}+U_{m-3,n+1}+U_{m,n-1}.\label{r-1}
\end{align}
This relations can be rewritten in the forms
\begin{align}
U_{m,n} &= (x-1)U_{m-1,n}-(y+1)U_{m-2,n}+(x^2-1-2y)U_{m-3,n}-(y+1)U_{m-4,n}\nn\\
&+ (x-1)U_{m-5,n}-U_{m-6,n}\nn\\
U_{m,n} &= (-x+y-1)U_{m,n-1}-(x^3-x+1-3xy-2y)U_{m,n-2}\label{r-2}\\
&+(-2x^3+x^2+2x-1+4xy+4y+y^2)U_{m,n-3}+(x^3-x+1-3xy-2y)U_{m,n-4}\nn\\
&-(-x+y-1)U_{m,n-5}+U_{m,n-6},\nn
\end{align}
in each of which changes only a single index. In contrast to the case of Chebyshev polynomials in
two variables associated with the Lie algebra $A_2$ \cite{DL}, the transition from relations
(\ref{r-1}) to (\ref{r-2}) is not obvious.

As it is shown in \cite{DKS}, this form is closely connected with the polynomials $P_1, P_2$,
standing in the denominator of the generating function. Namely, the polynomials $P_1, P_2$ are
the characteristic polynomials of matrices $M_1^{\pm}, M_2^{\pm}$. From here the
recurrence relation in the form (\ref{r-2}) immediately follows.

Let us represent the relations (\ref{r-2}) in a matrix form
\begin{equation*}
U_{m,n} = U_mM_x^nV_0,\quad U_{m,n} = U_nM_y^nV_0,
\end{equation*}
where $U_m$ and $U_n$ the following row-matrices
\begin{align*}
U_{m} &= (U_{5,n},-U_{4,n},U_{3,n},-U_{2,n},U_{1,n},U_{0,n}),\\
U_{n} &= (U_{m,5},-U_{m,4},U_{m,3},-U_{m,2},U_{m,1},U_{m,0});
\end{align*}
$V_0$ - column matrix
\begin{equation*}
V_{0} = (0,0,0,0,0,1)^T,
\end{equation*}
and matrices $M_x, M_y$ are of the following form
\begin{equation*}
M_x =\left(
             \begin{array}{cccccc}
               x-1 & 1 & 0 & 0 & 0& 0\\
               -(y+1) & 0 & 1 & 0 & 0 &0 \\
               x^2-1-2y & 0 & 0 & 1 & 0 &0\\
               -(y+1) & 0 & 0 & 0 & 1 &0\\
               x-1 & 0 & 0 & 0 & 0 &1\\
               -1&0&0&0&0&0\\
             \end{array}
           \right),
\end{equation*}
\begin{equation*}
M_y=\left(
             \begin{array}{cccccc}
               -x+y-1 & 1 & 0 & 0 & 0 &0\\
               -(x^3-x+1-3xy-2y) & 0 & 1 & 0 & 0 &0\\
               -2x^3+x^2+2x-1+4xy+4y+y^2 & 0 & 0 & 1 & 0 &0\\
              -(x^3-x+1-3xy-2y) & 0 & 0 & 0 & 1 &0\\
               -x+y-1 & 0 & 0 & 0 & 0 &1\\
               -1&0&0&0&0&0\\
             \end{array}
           \right).
\end{equation*}
The polynomials $P_1, P_2$ are the minimal polynomials for the matrices $M_x, M_y$.
Given above relations
\begin{equation*}
U_{m,n} = U_mM_x^nV_0,\quad U_{m,n} = U_nM_y^nV_0,
\end{equation*}
can be used for several other variants of receiving of  generating functions, and  also for
some other purposes, for example, to calculate polynomials with negative indices.
\medskip

{\bf Acknowledgments} This work is supported by RFBR under the grant 15-01-03148.

\end{document}